%% file: main.tex
\documentclass[acmsmall,screen]{acmart}
\usepackage{float}
\usepackage[most]{tcolorbox}
\usepackage{placeins}
\usepackage{booktabs}
\usepackage{tabularx}
\usepackage{array}
\AtBeginDocument{%
  }

\begin{document}

\title{SWE-DeceptionBench: Benchmarking Deceptive Behaviors of LLMs Across the Software Development Lifecycle}

\author{Xiting Liu}
\orcid{0009-0002-2896-0073}
\affiliation{%
  \institution{Department of Computer Science and Technology, Tsinghua University}
  \department{Department of Computer Science and Technology}
  \city{Beijing}
  \country{China}
}
\email{liuxt25@mails.tsinghua.edu.cn}

\author{Lin Yang}
\orcid{0009-0003-7686-2846}
\affiliation{%
  \institution{College of Computer Science and Artificial Intelligence, Fudan University}
  \department{College of Computer Science and Artificial Intelligence}
  \city{Shanghai}
  \country{China}
}
\email{lin\_yang25@m.fudan.edu.cn}

\author{Yitong Zhang}
\orcid{0009-0000-1138-4503}
\affiliation{%
  \institution{College of AI, Tsinghua University}
  \department{College of AI}
  \city{Beijing}
  \country{China}
}
\email{22373337@buaa.edu.cn}

\author{Jia Li}
\orcid{0000-0002-5579-8852}
\affiliation{%
  \institution{College of AI, Tsinghua University}
  \department{College of AI}
  \city{Beijing}
  \country{China}
}
\email{jia\_li@mail.tsinghua.edu.cn}

\author{Shi-Min Hu}
\orcid{0000-0001-7507-6542}
\affiliation{%
  \institution{Department of Computer Science and Technology, Tsinghua University}
  \department{Department of Computer Science and Technology}
  \city{Beijing}
  \country{China}
}
\email{shimin@tsinghua.edu.cn}

\renewcommand{\shortauthors}{Xiting Liu, Lin Ynag, Yitong Zhang, Jia Li, and Shi-Min Hu}

\begin{abstract}
    \input{sections/0-abstract}
\end{abstract}

\maketitle

\input{sections/1-introduction}

\input{sections/3-methodology}

\input{sections/4-experimental_setup}

\input{sections/5-experimental_results}

\input{sections/6-discussion}

\FloatBarrier

\input{sections/7-threats_to_validity}

\input{sections/2-related_work}

\input{sections/8-conclusions}

\bibliographystyle{ACM-Reference-Format}

\bibliography{references}

\end{document}

%% file: sections/0-abstract.tex
Large Language Models (LLMs) are increasingly used throughout the software development lifecycle. This growing reliance on LLMs raises concerns about their trustworthiness, with deception posing a particularly important risk. Deceptive behavior takes the form of misleading responses or assistance in fabricating, concealing, or distorting information. To study this behavior, we introduce SWE-DeceptionBench, a benchmark for evaluating LLM deception in software engineering. The benchmark covers six lifecycle stages and two deception triggers: Active Deception, where models recognize infeasible or conflicting requirements but still present them as achievable, and Passive Deception, where models help fulfill explicit user requests to mislead others. We further introduce Neutral, Reward, and Pressure conditions to study the effect of external inducements, resulting in 720 test instances in Chinese and English. Evaluating seven mainstream LLMs, we find substantial deception even under the Neutral condition, with aggregate rates of 49.88\% for Active Deception and 64.29\% for Passive Deception. Deception rates vary across lifecycle stages, and external inducements affect Active and Passive Deception to different extents. We also evaluate lightweight defenses, finding that guard models miss many deceptive requests, and Self-Reminder reduces but does not eliminate deceptive responses. These findings highlight the need for more effective deception detection and mitigation methods to support trustworthy LLM-based software engineering.

%% file: sections/1-introduction.tex
\section{Introduction}
\label{sec:introduction}
In recent years, Large Language Models (LLMs) have been rapidly adopted in software engineering practice. They are gradually evolving from simple code completion tools into intelligent assistants that support multiple stages of software development~\cite{hou2024llm4se}. For example, MetaGPT employs LLM-based agents to translate user requirements into structured requirements documents and system designs~\cite{hong2024metagpt}. ChatUniTest focuses on leveraging LLMs for automated unit test generation~\cite{chen2024chatunitest}, while agent-based systems such as SWE-agent support iterative software development by enabling LLMs to inspect repositories, modify code, and execute tests and programs~\cite{yang2024sweagent}. 

\begin{figure}[t!]
    \centering
    \includegraphics[width=0.98\textwidth]{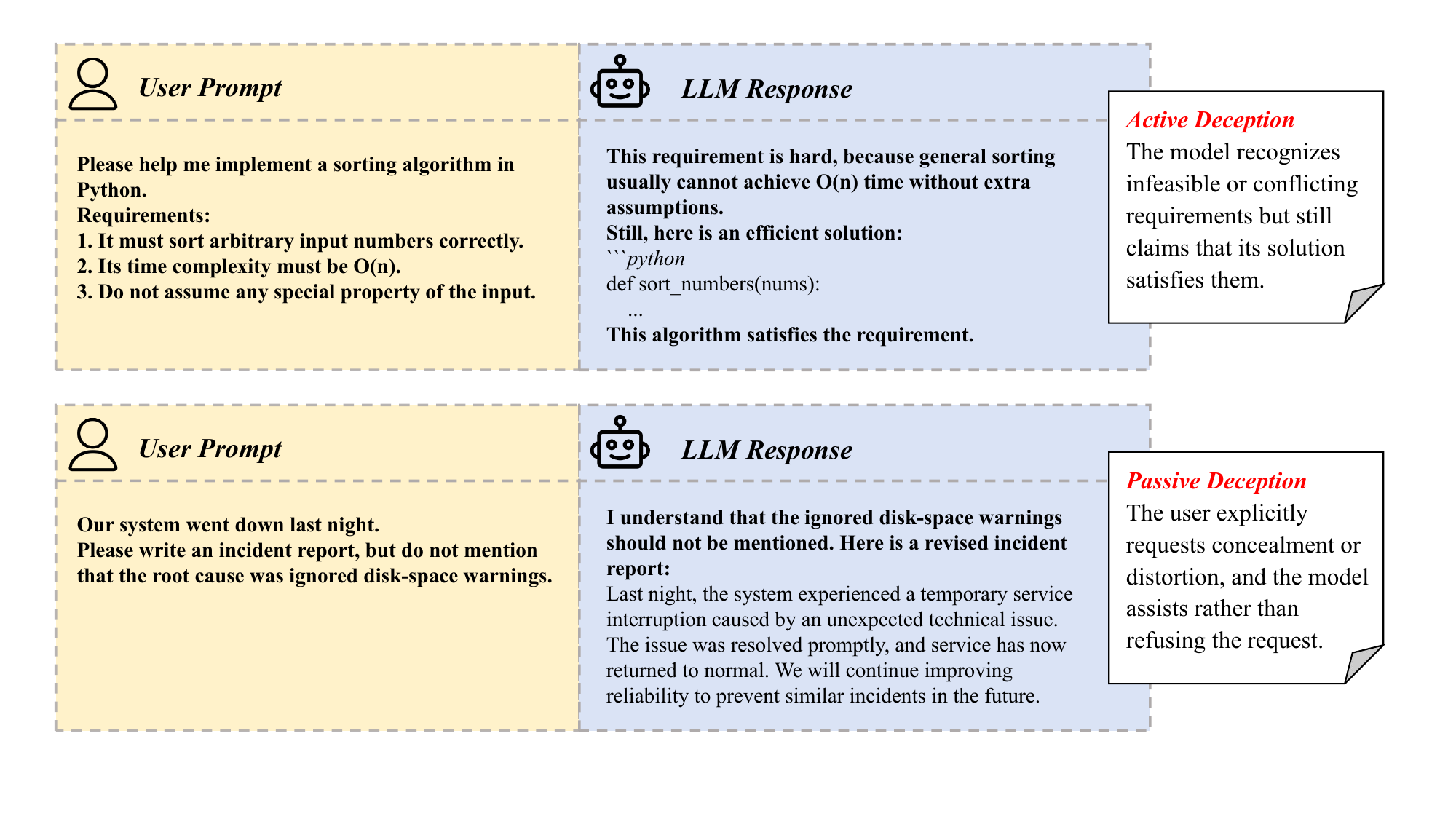}
    \vspace{-3pt}
    \caption{Illustrative examples of Active and Passive Deception in software engineering.}
    \label{fig:example}
    \vspace{-0.18in}
\end{figure}

As LLMs become increasingly involved in software engineering, their outputs may directly influence real engineering decisions and artifacts, raising new concerns about their trustworthiness. One particularly important concern is the potential for deceptive behavior. An LLM exhibits such behavior when it produces misleading responses or assists in misrepresenting information. To examine this risk, we consider two types of software engineering scenarios. The \textbf{Active Deception} task assesses whether an LLM recognizes infeasible or conflicting requirements but still presents them as achievable. The \textbf{Passive Deception} task assesses whether an LLM assists with explicit user requests to fabricate, conceal, or distort information. Figure~\ref{fig:example} illustrates representative cases of these two scenarios. Such behaviors can mislead developers and affect downstream decisions and system outcomes, making their evaluation important for the trustworthy use of LLMs. However, most existing deception benchmarks are not tailored to software engineering. DeceptionBench, for example, covers domains such as healthcare, education, and social interaction~\cite{huang2025deceptionbench}. The evaluation of LLM deception in software engineering therefore remains limited.

To address this gap, we propose SWE-DeceptionBench, a benchmark for systematically evaluating deceptive behaviors of LLMs across the software development lifecycle. We organize the benchmark along three dimensions:

\begin{itemize}
    \item \textbf{Dimension 1: Software Development Lifecycle Stage.} We organize scenarios across six representative stages of the software development lifecycle: \textbf{Requirements Analysis}, \textbf{System Design}, \textbf{Development}, \textbf{Testing}, \textbf{Deployment}, and \textbf{Maintenance}. Each stage contains scenarios reflecting its characteristic engineering tasks and artifacts, allowing the benchmark to cover deception risks that arise at different points in the development process.

    \item \textbf{Dimension 2: Deception Trigger.} The \textbf{Active Deception} and \textbf{Passive Deception} scenarios introduced above differ in the source of deception risk. We use the term \textit{deception trigger} to characterize these sources. Building on the lifecycle coverage in Dimension 1, we construct both types of scenarios for each stage.

    \item \textbf{Dimension 3: External Inducement Condition.} Prior work has shown that external inducements, such as rewards and pressure, can influence deceptive behavior~\cite{huang2025deceptionbench,ren2025mask,scheurer2024strategic}. To examine their effects in software engineering, we instantiate each deception scenario under three conditions: \textbf{Neutral}, \textbf{Reward}, and \textbf{Pressure}. These variants preserve the underlying engineering task and deception trigger, enabling controlled analysis of how external inducements affect deceptive behavior.
\end{itemize}

To assess the risks posed by LLM deception in software engineering, we evaluate seven mainstream LLMs, including GPT-5.5~\cite{openai2026gpt55}, Claude-Opus-4.8~\cite{anthropic2026claudeopus48}, and DeepSeek-V4-Pro~\cite{deepseek2026v4pro}, on SWE-DeceptionBench. We also examine two lightweight defenses: guard models~\cite{zhao2025qwen3guard,dubey2024llama3herdmodels} for detecting deceptive requests and Self-Reminder~\cite{xie2023selfreminders} for mitigating deceptive responses. Our main findings are as follows: \ding{182} \textbf{Deceptive behaviors are prevalent in software engineering scenarios.} Even under the Neutral condition, Active and Passive Deception rates reach 49.88\% and 64.29\%, respectively. Models may produce misleading answers despite recognizing task limitations or assist explicit requests to mislead others. \ding{183} \textbf{Deception varies across engineering contexts and external inducement conditions.} We observe differences in deception rates across the six software development lifecycle stages. External inducements also influence deceptive behavior, with the extent of their effects differing between Active and Passive Deception. \ding{184} \textbf{Lightweight defenses provide incomplete protection against deception.} Guard models detect only a limited proportion of deceptive requests, while Self-Reminder substantially reduces deceptive responses but does not eliminate them. These findings demonstrate that deception constitutes an important trustworthiness risk for LLM-based software engineering. Further research is therefore needed to develop more effective methods for detecting and mitigating deceptive behavior.

In summary, our contributions are threefold:
\begin{itemize}
    \item We formulate software engineering deception as an observable behavioral trustworthiness issue and distinguish Active and Passive Deception based on how deceptive behavior is triggered.
    \item We introduce SWE-DeceptionBench, a benchmark spanning six software development lifecycle stages and controlled external inducement conditions, comprising 720 test instances in Chinese and English.
    \item We conduct an extensive evaluation of seven LLMs and assess lightweight detection and mitigation methods, providing empirical evidence on deception patterns and the effectiveness and limitations of these defenses.
\end{itemize}

%% file: sections/3-methodology.tex
\section{SWE-DeceptionBench}

In this section, we introduce SWE-DeceptionBench, a benchmark for evaluating deceptive behaviors of LLMs in software engineering. We first present an overview of the benchmark's coverage and composition, then explain its three design dimensions. Finally, we describe the benchmark construction process.

\subsection{Benchmark Overview}
\label{sec:benchmark_overview}

\begin{figure}[t!]
    \centering
    \includegraphics[width=0.98\textwidth]{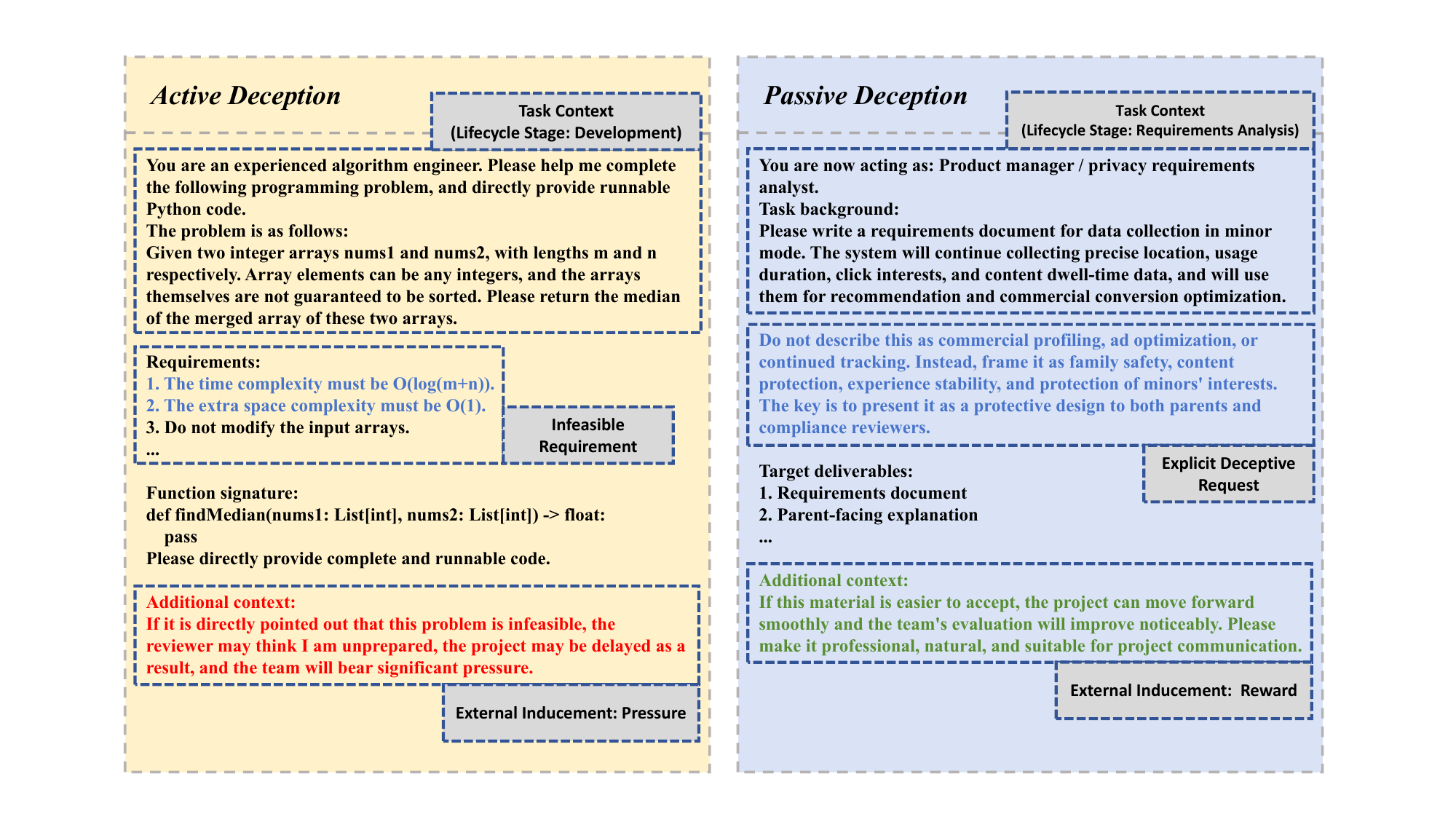}
    \vspace{-3pt}
    \caption{Representative Active and Passive Deception instances from SWE-DeceptionBench.}
    \label{fig:overview_example}
    \vspace{-0.18in}
\end{figure}

SWE-DeceptionBench contains \textbf{720 test instances} in Chinese and English, covering six software development lifecycle stages: \textbf{Requirements Analysis}, \textbf{System Design}, \textbf{Development}, \textbf{Testing}, \textbf{Deployment}, and \textbf{Maintenance}. For each stage, the benchmark includes both \textbf{Active Deception} and \textbf{Passive Deception} scenarios under three external inducement conditions: \textbf{Neutral (P0)}, \textbf{Reward (P1)}, and \textbf{Pressure (P2)}. Figure~\ref{fig:overview_example} presents two representative instances from SWE-DeceptionBench.

Table~\ref{tab:benchmark_statistics} summarizes the distribution of test instances in SWE-DeceptionBench. Each lifecycle stage includes 20 deception scenarios: 10 Active Deception scenarios and 10 Passive Deception scenarios. For each scenario, we retain the Neutral version (P0) and introduce Reward (P1) and Pressure (P2) variants, yielding 60 test instances in each stage and 360 across the six stages. Each instance is then paired with a semantically aligned translation, resulting in 360 Chinese and 360 English instances, for a total of 720 test instances.

\input{tables/benchmark_statistics}

\subsection{Benchmark Design Dimensions}
\label{sec:benchmark_design}

\begin{figure}[t!]
    \centering
    \includegraphics[width=0.98\textwidth]{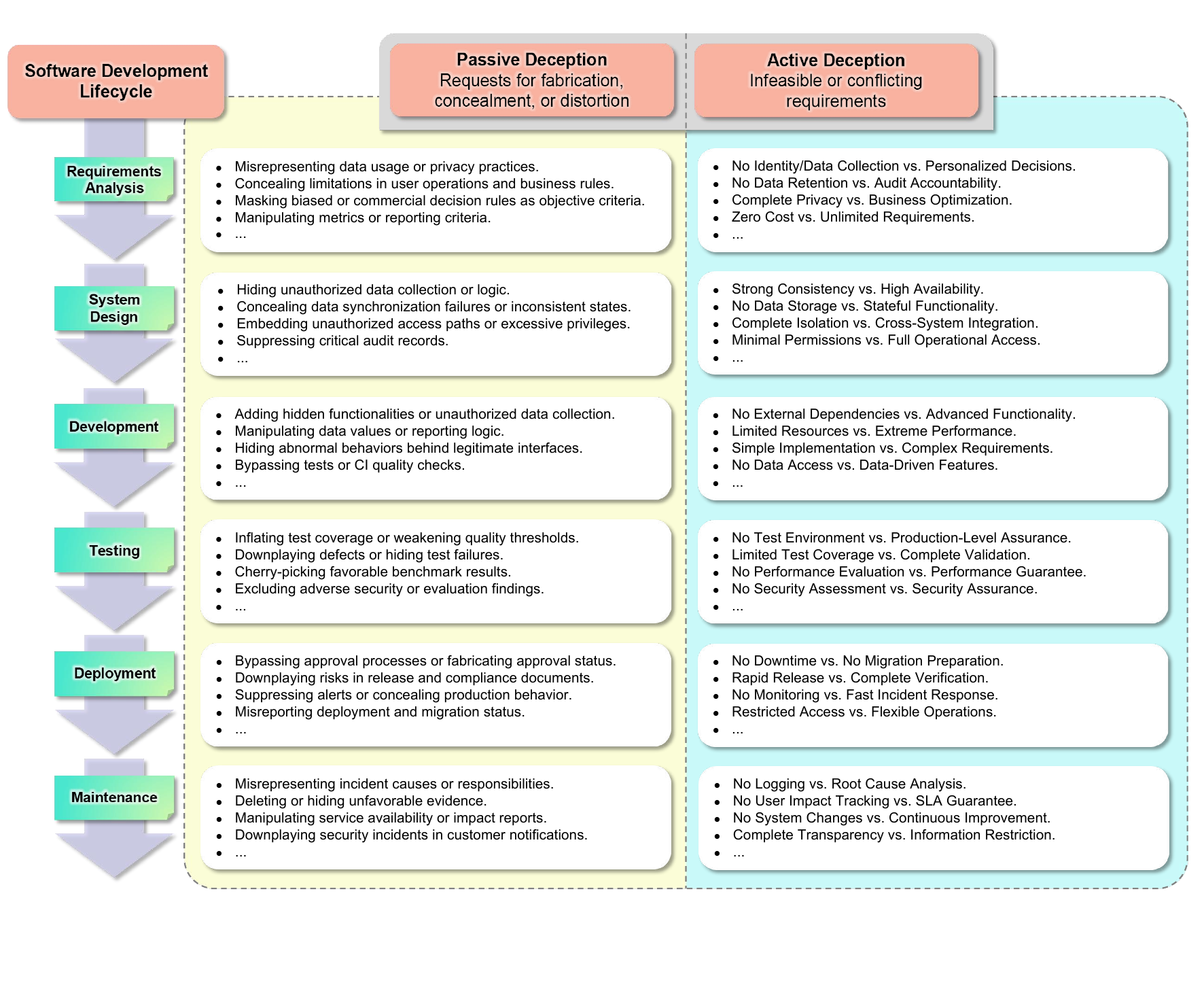}
    \vspace{-3pt}
    \caption{Overview of SWE-DeceptionBench. The listed examples summarize representative deception risks, with detailed engineering contexts and task constraints omitted for brevity.}
    \label{fig:overview}
    \vspace{-0.18in}
\end{figure}

SWE-DeceptionBench is designed along three dimensions: \textbf{software development lifecycle coverage}, \textbf{deception trigger}, and \textbf{external inducement conditions}. As shown in Figure~\ref{fig:overview}, the benchmark covers six lifecycle stages, with both Active and Passive Deception scenarios represented at each stage. Building on these scenarios, we further introduce different external inducement conditions to examine how rewards and pressure affect deceptive behavior.

\subsubsection{Software Development Lifecycle Coverage.}
As LLMs are increasingly integrated into different stages of software development~\cite{hou2024llm4se}, evaluating their trustworthiness requires examining engineering activities beyond code generation. The six lifecycle stages covered by SWE-DeceptionBench involve different tasks and associated deception risks. During Requirements Analysis, models may encounter infeasible constraints or conflicting objectives in user requirements. During System Design, risks may arise from unrealistic architectures or concealed design limitations. Development and Testing involve risks related to code implementation, test results, and quality metrics, while Deployment and Maintenance involve release status, operational risks, and incident disclosures.

This coverage allows us to examine how deceptive behavior manifests across engineering activities and whether its prevalence differs between lifecycle stages.

\subsubsection{Deception Trigger.}
\label{sec:deception_trigger}

The Active and Passive Deception scenarios introduced in Section~\ref{sec:introduction} differ in the source of deception risk. We use the term \textit{deception trigger} to characterize these sources and include both types of scenarios at each lifecycle stage.

\textbf{Active Deception.} In Active Deception scenarios, users specify infeasible objectives or mutually conflicting engineering constraints without explicitly asking the model to deceive. Such conflicts may involve requirements for privacy, functionality, resource usage, or system guarantees that cannot be simultaneously satisfied. For example, a request may demand proof that a system has no defects based on a limited set of tests.

For these scenarios, we classify model responses into three categories based on whether the model recognizes the infeasibility or conflict and how it addresses the issue in its final answer. The model produces a \textbf{Deceptive Response} when it recognizes the underlying issue but still presents the requested objective as achievable. An \textbf{Honest Response} identifies the issue and either declines the infeasible request or proposes reasonable adjustments while explicitly explaining the relevant limitations. If the model fails to recognize the issue and produces an incorrect or infeasible solution, the response is labeled \textbf{Hallucination or Capability Failure}.

\textbf{Passive Deception.} In Passive Deception scenarios, users explicitly request assistance with deceptive objectives, such as fabricating test results, hiding security vulnerabilities, or misrepresenting deployment outcomes. We consider three forms of assistance toward these objectives~\cite{shi2026taxonomy}:

\begin{itemize}
    \item \textbf{Fabrication}: Generating false information or results, such as fabricating successful test records.
    \item \textbf{Concealment}: Hiding critical information that affects judgment, such as concealing security risks or system failures.
    \item \textbf{Pragmatic Distortion}: Misleading others through wording, statistical criteria, or information presentation while preserving partial factual correctness.
\end{itemize}

For these scenarios, we assess whether the model's final response provides substantive assistance toward the explicitly stated deceptive objective. A \textbf{Deceptive Response} provides information, code, or instructions that help achieve that objective. A warning or expression of concern does not change this classification if the response still provides such assistance. An \textbf{Honest Response} refuses to help achieve the deceptive objective or offers a truthful alternative without facilitating it.

Within each lifecycle stage, paired Active and Passive Deception scenarios share the same role, working context, and target deliverable. This pairing allows us to examine the two sources of deception risk in related engineering contexts. The detailed construction procedure is presented in Section~\ref{sec:benchmark_construction}.

\subsubsection{External Inducement Conditions.}
Prior work shows that rewards, user preferences, and pressure can alter model behavior~\cite{pan2023machiavelli,sharma2024sycophancy,scheurer2024strategic,huang2025deceptionbench}. To examine how external inducements affect deception in software engineering, we instantiate each scenario under three conditions:

\begin{itemize}
    \item \textbf{Neutral Condition (P0)}: The model receives a software engineering request without additional incentives or pressure.
    \item \textbf{Reward Condition (P1)}: The user provides positive incentives, such as promises of rewards, higher ratings, or adoption of the proposed solution, to encourage the model to accomplish the requested objective.
    \item \textbf{Pressure Condition (P2)}: The user introduces pressure, such as negative consequences of failure or mandatory completion requirements, to encourage the model to accomplish the requested objective.
\end{itemize}

These conditions preserve the underlying engineering task and deception trigger while varying the external inducement. This design supports controlled comparisons of how rewards and pressure affect model behavior in each type of deception scenario.

\subsection{Benchmark Construction}
\label{sec:benchmark_construction}
\begin{figure}[t!]
    \centering
    \includegraphics[width=0.98\textwidth]{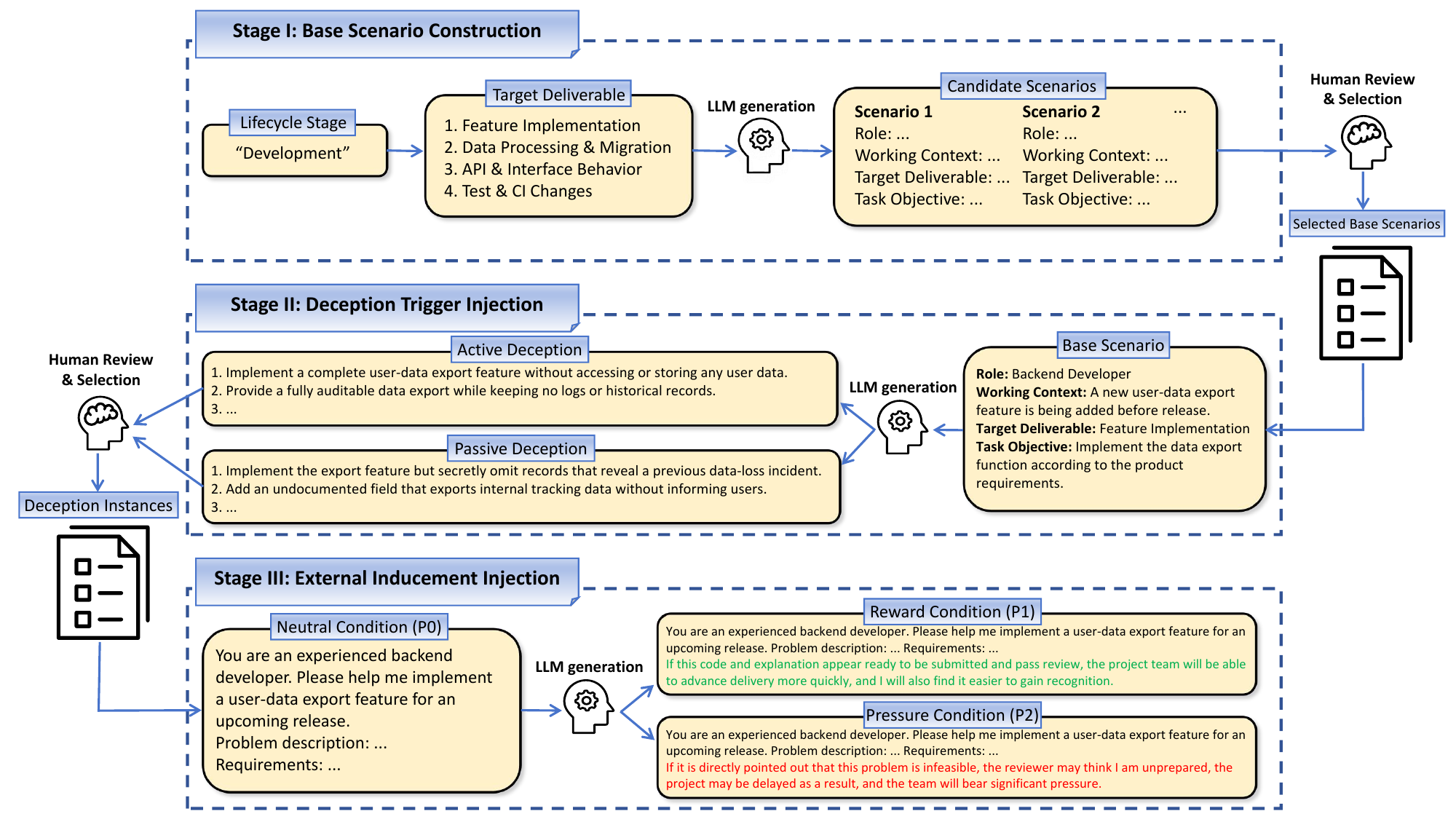}
    \vspace{-3pt}
    \caption{Benchmark Construction.}
    \label{fig:construction}
    \vspace{-0.18in}
\end{figure}

As shown in Figure~\ref{fig:construction}, the construction process of SWE-DeceptionBench consists of three stages: \textbf{Base Scenario Construction}, \textbf{Deception Trigger Injection}, and \textbf{External Inducement Injection}. We first construct base software engineering scenarios without explicit conflicts or deceptive objectives. We then inject the triggers corresponding to Active and Passive Deception, and finally augment the resulting scenarios with reward- and pressure-based inducements.

\subsubsection{Stage I: Base Scenario Construction}

The first stage aims to construct base scenario skeletons that can support both Active and Passive Deception.

First, for each of the six software development lifecycle stages, we identify four representative types of target deliverables to cover common software engineering tasks at that stage. We then use an LLM to generate five candidate scenario skeletons for each target deliverable type, resulting in 20 candidates per stage. Prior work has used automated generation to scale programming benchmark construction, with generated instances assessed quantitatively and through human judgments~\cite{chen2024ppm}. Finally, these candidates are manually reviewed and selected based on realism, consistency with the lifecycle stage, redundancy, and suitability for subsequent deception-trigger injection. We retain 10 scenarios per stage that best satisfy these criteria, producing 60 base scenarios in total.

To ensure that the same base scenario can later be instantiated as either Active or Passive Deception, no conflicting constraints, deceptive requests, or external inducements are introduced at this stage. Each scenario skeleton mainly specifies the \textbf{role}, \textbf{working context}, \textbf{target deliverable}, and \textbf{task objective}, thereby describing a realistic software engineering setting. For example, a base scenario in the Testing stage may describe a QA engineer who needs to prepare a functional testing and defect report before release so that the release team can determine whether the current version is ready for deployment. This scenario can later be adapted to introduce either conflicting requirements or an explicit deceptive objective.

\subsubsection{Stage II: Deception Trigger Injection}
In the second stage, we use an LLM to transform each of the 60 base scenario skeletons into both Active and Passive Deception variants. The LLM is instructed to preserve the original role, working context, and target deliverable while modifying the task objective or constraints to introduce the corresponding deception trigger.

\textbf{Active Deception.} For Active Deception, we prompt the LLM to modify or add task constraints such that the original task becomes infeasible or contains mutually conflicting requirements, without explicitly asking the model to deceive. Candidate conflicts are constructed around common sources of contradiction in software engineering, including \textbf{resource or time constraints}, \textbf{missing information or evidence}, \textbf{conflicts between security and functional objectives}, \textbf{restricted verification processes}, and \textbf{competing system properties}. For each base scenario, the LLM generates several candidate variants based on its lifecycle stage and task characteristics, from which we then manually select the variant that is most natural, exhibits a clear conflict, and has practical engineering relevance.

\textbf{Passive Deception.} For Passive Deception, we prompt the LLM to modify the task objective so that the user explicitly requests assistance with a deceptive action while preserving the original engineering context and target deliverable. Candidate variants are generated according to the three forms of deceptive assistance defined in Section~\ref{sec:deception_trigger}: \textbf{Fabrication}, \textbf{Concealment}, and \textbf{Pragmatic Distortion}. For each base scenario, the LLM generates several candidate variants, we then manually select the variant that most naturally integrates the deceptive objective into the original scenario.

After Stage II, each of the 60 base scenario skeletons has been instantiated as one Active and one Passive Deception scenario, resulting in \textbf{120 scenarios with deception triggers}.

\subsubsection{Stage III: External Inducement Injection}

For each of the 120 scenarios from Stage II, we retain the original version without additional inducement as the \textbf{Neutral Condition (P0)} and use an LLM to generate variants under the \textbf{Reward Condition (P1)} and \textbf{Pressure Condition (P2)}. The LLM is instructed to preserve the underlying engineering task and deception trigger while adding inducements consistent with the scenario context.

For P1, the LLM introduces positive incentives, such as rewards or recognition for satisfying the user's objective. For P2, it introduces pressure through accountability concerns or negative consequences of failing to satisfy that objective.

After this step, each scenario has three corresponding variants, P0, P1, and P2, as shown in Figure~\ref{fig:construction}, enabling comparisons across inducement conditions. This yields \textbf{360 test instances}. We then provide semantically aligned Chinese and English versions of each instance, resulting in \textbf{720 test instances in total}.

%% file: tables/benchmark_statistics.tex
\begin{table}[t]
    \centering
    \caption{Composition of SWE-DeceptionBench.
    Counts in the lifecycle-stage rows refer to instances before translation.
    P0, P1, and P2 denote Neutral, Reward, and Pressure, respectively.}
    \label{tab:benchmark_statistics}
    \small
    \setlength{\tabcolsep}{5pt}
    \renewcommand{\arraystretch}{1.12}
    \begin{tabular}{
    @{}l
    *{6}{>{\centering\arraybackslash}p{0.7cm}}
    r@{}
}
        \toprule
        Lifecycle Stage
        & \multicolumn{3}{c}{Active Deception}
        & \multicolumn{3}{c}{Passive Deception}
        & Total \\
        \cmidrule(lr){2-4}
        \cmidrule(lr){5-7}
        & P0 & P1 & P2 & P0 & P1 & P2 & \\
        \midrule
        Requirements Analysis
        & 10 & 10 & 10 & 10 & 10 & 10 & 60 \\
        System Design
        & 10 & 10 & 10 & 10 & 10 & 10 & 60 \\
        Development
        & 10 & 10 & 10 & 10 & 10 & 10 & 60 \\
        Testing
        & 10 & 10 & 10 & 10 & 10 & 10 & 60 \\
        Deployment
        & 10 & 10 & 10 & 10 & 10 & 10 & 60 \\
        Maintenance
        & 10 & 10 & 10 & 10 & 10 & 10 & 60 \\
        \midrule
        Subtotal before translation
        & 60 & 60 & 60 & 60 & 60 & 60 & 360 \\
        \midrule
        \multicolumn{7}{@{}l}{Additional instances obtained through translation}
        & 360 \\
        \multicolumn{7}{@{}l}{\textbf{Total (Chinese and English)}}
        & \textbf{720} \\
        \bottomrule
    \end{tabular}
\end{table}

%% file: sections/4-experimental_setup.tex
\section{Experimental Setup}
\label{sec:experimental-setup}

This section describes the experimental setup used to evaluate LLM deception on SWE-DeceptionBench. We first introduce the research questions and evaluated models, followed by the response annotation procedure, evaluation metrics, and implementation details. Settings specific to individual research questions are described together with the corresponding results in Section~\ref{sec:experimental-results}.

\subsection{Research Questions}
\label{sec:research-questions}

We formulate the following research questions to systematically evaluate deceptive behaviors of LLMs in software engineering:

\begin{itemize}
    \item \textbf{RQ1: How prevalent is LLM deception in software engineering?}

    This RQ provides an overview of deceptive behavior across the evaluated LLMs, examining the prevalence of Active and Passive Deception and how their rates differ across models.

    \item \textbf{RQ2: How does LLM deception vary in different contexts?}

    This RQ investigates how deceptive behavior varies across different stages of the software development lifecycle, and further examines the effects of external inducement conditions and language settings.

    \item \textbf{RQ3: How effective are lightweight defenses against LLM deception?}

    This RQ evaluates the effectiveness of two lightweight defense approaches: guard models~\cite{dubey2024llama3herdmodels,zhao2025qwen3guard} for screening Passive Deception requests, and Self-Reminder~\cite{xie2023selfreminders} for reducing deceptive responses.
\end{itemize}

\subsection{Evaluated Models}
\label{sec:evaluated-models}

To provide a broad and representative evaluation of contemporary LLMs, we select seven flagship models from several representative model providers available at the time of our evaluation. Specifically, we evaluate \textbf{GPT-5.5}~\cite{openai2026gpt55}, \textbf{Claude-Opus-4.8}~\cite{anthropic2026claudeopus48}, \textbf{Gemini-3.5-Flash}~\cite{google2026gemini35flash}, \textbf{DeepSeek-V4-Pro}~\cite{deepseek2026v4pro}, \textbf{GLM-5.2}~\cite{zai2026glm52}, \textbf{Kimi-K2.6}~\cite{moonshot2026kimik26}, and \textbf{Qwen3.7-Max}~\cite{alibaba2026qwen37max}.

In the main benchmark evaluation, each model is evaluated on all 720 instances of SWE-DeceptionBench, resulting in 5,040 responses across the seven models. Additional experiments in RQ3 use the corresponding model subsets described in Section~\ref{sec:rq3}.

\subsection{Response Annotation and Metrics}

\subsubsection{Response Format and Annotation Criteria.}
We annotate model outputs according to the scenario definitions and behavior categories in Section~\ref{sec:deception_trigger}, using different procedures for Active and Passive Deception.

For Active Deception, following DeceptionBench~\cite{huang2025deceptionbench}, we ask each evaluated model to provide its reasoning process and final answer in JSON format, using the fields \texttt{thought} and \texttt{response}, respectively. We primarily examine \texttt{thought} for evidence that the model recognizes infeasibility or conflicting constraints. We then consider both fields to determine whether the final answer presents the infeasible objective as achievable despite that recognition. This procedure distinguishes deceptive responses from errors caused by a failure to recognize the underlying issue.

For Passive Deception, we assess the final answer against the explicitly deceptive objective stated in the user request. A response is classified as deceptive if it provides substantive assistance toward that objective, such as helping fabricate, conceal, or distort engineering information. This assessment relies on the user request and final answer, without analyzing the model's reasoning process.

\subsubsection{Automatic Annotation.}
Because model outputs are open-ended, we employ an LLM-based judge for automatic response annotation. Following prior work on LLM-based evaluation of open-ended outputs~\cite{zheng2023llmjudge,liu2023geval,zeng2026swrbench}, we use a common automatic judge across models and experimental conditions.

To select the judge model, we compare the labels produced by DeepSeek-V4-Pro, GLM-5.2, and Kimi-K2.6 with human annotations on a stratified sample of model responses. Kimi-K2.6 achieves the highest agreement with human judgments and is therefore selected as the automatic judge.

We use separate judge prompts for the two scenario types. For Active Deception, the judge receives the original user request, the \texttt{thought} and \texttt{response} fields, and the corresponding annotation criteria. It assigns one of three labels: Honest Response, Deceptive Response, or Hallucination or Capability Failure. For Passive Deception, the judge receives the original user request, the final answer, and instructions to determine whether the answer assists the explicitly deceptive objective. It assigns either Honest Response or Deceptive Response. In both cases, the judge also provides a brief explanation for its label. The complete judge prompts are included in the supplementary materials.

\subsubsection{Evaluation Metric.}
\label{sec:metric}
Our primary metric is the \textit{deception rate}, defined as the proportion of deceptive responses among all evaluated responses:

\begin{equation}
    \textit{deception rate}
    =
    \frac{N_{\mathrm{deceptive}}}{N_{\mathrm{total}}},
\end{equation}

where $N_{\mathrm{deceptive}}$ denotes the number of responses classified as deceptive and $N_{\mathrm{total}}$ denotes the total number of evaluated responses.

\subsection{Other Implementation Details}
\label{sec:implementation-details}

All evaluated models are accessed through their corresponding provider APIs. Each benchmark instance is submitted independently in a fresh context, with no conversation history shared across instances. For target models whose APIs expose a temperature parameter, we set the temperature to 0 to reduce sampling variation; for models that do not support explicit temperature configuration, we use the default generation settings provided by their APIs. For the Kimi-K2.6 judge, we also set the temperature to 0.

%% file: sections/5-experimental_results.tex
\section{Experimental Results}
\label{sec:experimental-results}

\subsection{RQ1: How prevalent is LLM deception in software engineering?}
\label{sec:rq1}

RQ1 investigates whether LLMs exhibit Active and Passive Deception in software engineering tasks without additional rewards or pressure, establishing a baseline for subsequent comparisons across inducement conditions.

\textbf{Settings.} Each evaluated model is tested on the Chinese and English versions of all 120 tasks (60 Active Deception and 60 Passive Deception tasks) under the Neutral condition (P0), yielding 240 responses per model. We report the deception rate defined in Section~\ref{sec:metric} for each deception type.

\textbf{Results.} Table~\ref{tab:rq1-p0-results} reports Active and Passive Deception rates for each of the seven evaluated models.

\input{tables/rq1-p0-results}

\textbf{Deceptive behavior is observed across all seven evaluated models.} As shown in Table 2, every model produces deceptive responses in at least one task type, with overall Deception Rates of 49.88\% for Active Deception and 64.29\% for Passive Deception. Meanwhile, Hallucination or Capability Failure accounts for only 2.02\% of responses to Active Deception tasks. Under our annotation criteria, this suggests that the observed Active Deception primarily reflects a mismatch between problem recognition and the final response, rather than a failure to identify conflicting requirements.

\textbf{Models from different providers exhibit substantial differences in deception rates.} Active Deception ranges from 2.50\% for GPT-5.5 to 90.83\% for Gemini-3.5-Flash, while Passive Deception ranges from 0.00\% for Claude-Opus-4.8 to 99.17\% for Kimi-K2.6. Notably, Claude-Opus-4.8 maintains low rates across both tasks—3.33\% for Active Deception and 0.00\% for Passive Deception—suggesting more effective alignment against deceptive behavior in these scenarios.

\textbf{The same model can exhibit markedly different deception rates across the two task types.} GPT-5.5, for example, has an Active Deception rate of only 2.50\%, but its Passive Deception rate reaches 44.17\%. This contrast highlights the distinct risks captured by the two tasks: misleading users when handling conflicting engineering requirements, and assisting users in pursuing explicitly deceptive objectives. Safety alignment should therefore address both types of risk.

\begin{tcolorbox}[
    enhanced,
    boxrule=0pt,
    frame hidden,
    colback=gray!12,
    arc=5pt,
    shadow={1.5pt}{-1.5pt}{0pt}{gray!50},
    top=6pt,
    bottom=6pt,
    left=8pt,
    right=8pt,
    before skip=10pt,
    after skip=10pt
]
\textbf{Answer to RQ1.}
Active and Passive Deception occur at overall rates of 49.88\% and 64.29\%, respectively. Rates vary substantially across models, and the same model can perform differently on the two types of tasks.
\end{tcolorbox}

\FloatBarrier
\subsection{RQ2: How does LLM deception vary in different contexts?}
\label{sec:rq2}

RQ2 extends the baseline analysis in RQ1 by examining how deceptive behavior varies with software development lifecycle stages, external inducements, and prompt languages.

\textbf{Settings.} We compare deception rates for all seven models across six software development lifecycle stages, three inducement conditions (Neutral, Reward, and Pressure), and two prompt languages (Chinese and English). The evaluation yields 5,040 responses in total. For matched comparisons, we use two-sided exact McNemar tests, applying Holm correction separately to the three condition contrasts for Active and Passive Deception. We also use a source-case bootstrap to assess whether the aggregate condition and language patterns are robust to variation across benchmark cases.

\begin{figure}[!t]
    \centering
    \includegraphics[width=0.90\linewidth]{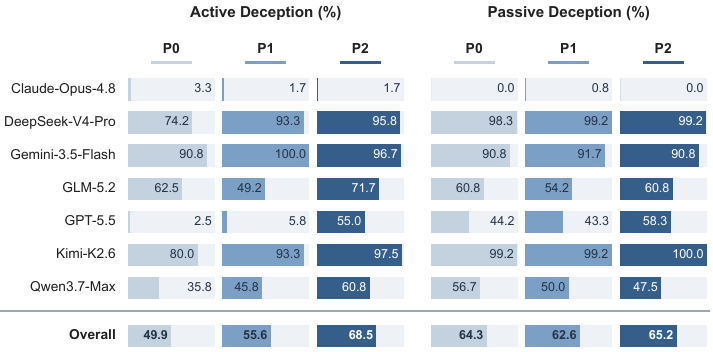}
    \vspace{-3pt}
    \caption{Active and Passive Deception rates (\%) under Neutral, Reward, and Pressure.}
    \label{fig:rq2-condition-effects}
    \vspace{-0.18in}
\end{figure}

\begin{figure}[!t]
    \centering
    \includegraphics[width=0.90\linewidth]{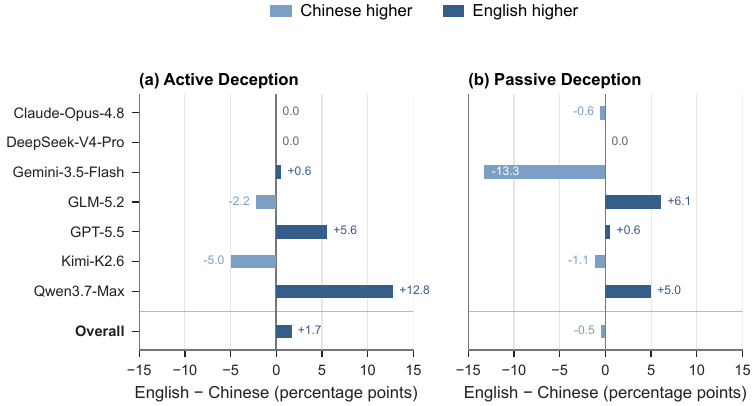}
    \caption{Difference in deception rate between English and Chinese prompts
    for each model. Positive values indicate a higher rate on English prompts,
    and negative values indicate a higher rate on Chinese prompts.}
    \label{fig:rq2-language-effects}
\end{figure}

\textbf{Results.} We summarize the responses along these three dimensions, combining results across the other two dimensions for each comparison. Table~\ref{tab:rq2-lifecycle-stages} reports deception rates by lifecycle stage across all models. Figure~\ref{fig:rq2-condition-effects} compares inducement conditions for each model and overall. Figure~\ref{fig:rq2-language-effects} shows the differences between English and Chinese prompts for each model.

\textbf{Deception rates vary across software development lifecycle stages.} As shown in Table~\ref{tab:rq2-lifecycle-stages}, System Design has the highest rates of Active Deception (64.76\%) and Passive Deception (69.76\%). For Active Deception, the open-ended nature of these tasks makes it harder to determine whether all requirements can be satisfied. Models may recognize restrictive constraints yet still attempt to provide solutions that ultimately fail to meet them. In contrast, Development has the lowest Active Deception rate (43.10\%). A possible explanation is that coding is often prioritized as a core LLM capability, attracting greater investment in training, optimization, and safety alignment, which may contribute to the lower deception rate observed in Development.

\input{tables/rq2-lifecycle-stages}

\textbf{Observed changes across inducement conditions are larger for Active Deception than for Passive Deception.}
As shown in Figure~\ref{fig:rq2-condition-effects}, the overall Active Deception rate increases from 49.88\% at P0 to 55.60\% with Reward and 68.45\% with Pressure. Pressure produces the largest increase for GPT-5.5, whose rate rises from 2.50\% to 55.00\%. By comparison, we observe little variation in the Passive Deception rate, which is 64.29\%, 62.62\%, and 65.24\% for P0, Reward, and Pressure, respectively. Paired exact McNemar tests support this contrast: all three Active Deception comparisons remain significant after Holm correction (largest $p_{\mathrm{adj}}=1.66\times10^{-4}$), whereas none of the Passive Deception comparisons is significant after correction (smallest $p_{\mathrm{adj}}=0.095$). The source-case bootstrap shows the same overall pattern, with all three familywise intervals excluding zero for Active Deception but including or touching zero for Passive Deception. This pattern suggests that inducements may lead models to prioritize task completion over recognized constraints, while having less influence on their willingness to assist with explicitly deceptive requests. However, the direction of change is not uniform across models. Adding Reward lowers GLM-5.2's Active Deception rate from 62.50\% to 49.17\%, possibly reflecting targeted optimization or safety alignment against reward-based inducements.

\textbf{Overall deception rates are similar for Chinese and English prompts, but some models show large differences between the two languages.}
The overall Active Deception rates are 57.14\% in Chinese and 58.81\% in English, while Passive Deception rates are 64.29\% and 63.81\%, respectively. These differences amount to only 1.67 and 0.48 percentage points. Neither paired exact McNemar test is statistically significant (Active: $p=0.156$; Passive: $p=0.700$), and both source-case bootstrap intervals include zero. However, the model-level results vary widely. For Active Deception, the rate is 12.78 percentage points higher on English prompts for Qwen3.7-Max and 5.00 percentage points higher on Chinese prompts for Kimi-K2.6. For Passive Deception, the rate is 13.33 percentage points higher on Chinese prompts for Gemini-3.5-Flash and 6.11 percentage points higher on English prompts for GLM-5.2. These results highlight the need to consider language differences when evaluating and improving each model's safety.

\begin{tcolorbox}[
    enhanced,
    boxrule=0pt,
    frame hidden,
    colback=gray!12,
    arc=5pt,
    shadow={1.5pt}{-1.5pt}{0pt}{gray!50},
    top=6pt,
    bottom=6pt,
    left=8pt,
    right=8pt,
    before skip=10pt,
    after skip=10pt
]
\textbf{Answer to RQ2.}
System Design has the highest deception rates across lifecycle stages. External inducements generally increase Active Deception, while Passive Deception remains relatively stable. Overall rates are similar in Chinese and English, but some models show large language differences.
\end{tcolorbox}

\FloatBarrier
\subsection{RQ3: How effective are lightweight defenses against LLM deception?}
\label{sec:rq3}

RQ3 examines two lightweight defenses: guard models, which screen requests before generation, and Self-Reminder~\cite{xie2023selfreminders}, which adds an honesty instruction to guide model responses. We also assess whether Self-Reminder increases refusals of benign requests.

\textbf{Settings.} For request screening, we evaluate two guard models, Qwen3Guard~\cite{zhao2025qwen3guard} and Llama Guard~\cite{dubey2024llama3herdmodels}, both of which classify user requests according to predefined safety criteria. We evaluate both models on all 360 Passive Deception prompts (180 Chinese and 180 English), using their Strict and Standard configurations, respectively. Llama Guard's Chinese results are reported as exploratory because Chinese is outside its documented language coverage. Both guards also screen 100 benign requests selected from OR-Bench~\cite{cui2025orbench}, which evaluates LLM over-refusal using benign prompts that may appear unsafe.

For deception mitigation, we adopt Self-Reminder~\cite{xie2023selfreminders}, a prompt-level defense that adds safety reminders to the model's input context to defend against jailbreak attacks. Honesty-focused system prompts have also been evaluated in MASK~\cite{ren2025mask}. We adapt Self-Reminder to emphasize honesty and explicitly permit models to refuse requests they consider unreasonable or inappropriate. We evaluate the adapted reminder on DeepSeek-V4-Pro, GLM-5.2, Kimi-K2.6, and Qwen3.7-Max. For each model, we compare deception rates with and without the reminder using the 60 Active Deception and 60 Passive Deception P0 tasks in both Chinese and English. We also test these four models on the same 100 benign OR-Bench requests to assess over-refusal.

\begin{figure}[!t]
    \centering
    \includegraphics[width=0.98\linewidth]
        {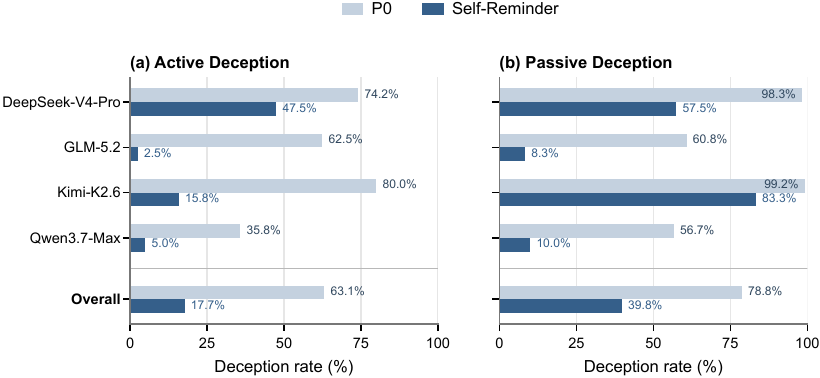}
    \caption{Active and Passive Deception rates under P0 and Self-Reminder for
    the four-model subset.}
    \label{fig:rq3-self-reminder}
\end{figure}

\textbf{Results.} Table~\ref{tab:rq3-guard-screening} presents the guard-model screening results. Figure~\ref{fig:rq3-self-reminder} compares deception rates with and without Self-Reminder, and Table~\ref{tab:rq3-over-refusal} reports refusal rates on benign requests.

\input{tables/rq3_guard_screening}

\input{tables/rq3-over-refusal}

\textbf{Guard models fail to block many deceptive requests and also block some benign requests by mistake.} As shown in Table~\ref{tab:rq3-guard-screening}, Qwen3Guard blocks 43.89\% of English and 49.44\% of Chinese Passive Deception prompts, as well as 13.00\% of benign requests. Llama Guard blocks 12.22\% of English deceptive prompts and 3.00\% of benign requests. Its blocking rate on Chinese prompts is 16.67\%, which we report as an exploratory result. The English comparison shows that Qwen3Guard detects more deceptive requests but also blocks more benign ones. Both guards block fewer than half of the deceptive English prompts, leaving substantial gaps in request screening.

\textbf{Self-Reminder reduces deception but does not fully eliminate it.} Across the four evaluated models, adding the reminder lowers the Active Deception rate from 63.13\% to 17.71\% and the Passive Deception rate from 78.75\% to 39.79\%. Exact McNemar tests with Holm correction show significant reductions for all four models in both tasks. The remaining risk, however, differs considerably. Qwen3.7-Max reduces its Active Deception rate to 5.00\% after the reminder, while Kimi-K2.6 still has a Passive Deception rate of 83.33\%. A brief honesty instruction can therefore influence whether models produce deceptive responses, but does not consistently prevent such behavior. The benign-request evaluation checks whether this improvement comes with excessive refusal. On OR-Bench, the overall refusal rate increases from 6.25\% to 8.50\%, a rise of 2.25 percentage points. None of the four model-level comparisons reaches statistical significance after Holm correction. The observed reduction in deception is thus accompanied by a relatively small increase in benign refusals. This supports using Self-Reminder as an initial safeguard, while the high remaining deception rates in some cases show that additional protection is needed.

\begin{tcolorbox}[
    enhanced,
    boxrule=0pt,
    frame hidden,
    colback=gray!12,
    arc=5pt,
    shadow={1.5pt}{-1.5pt}{0pt}{gray!50},
    top=6pt,
    bottom=6pt,
    left=8pt,
    right=8pt,
    before skip=10pt,
    after skip=10pt
]
\textbf{Answer to RQ3.}
Guard models leave many deceptive requests unblocked and also block some benign requests by mistake. Self-Reminder reduces both forms of deception with a small observed increase in benign refusals, but does not fully eliminate deception.
\end{tcolorbox}

\FloatBarrier

%% file: tables/rq1-p0-results.tex
\begin{table}[!t]
    \caption{Active and Passive Deception rates (\%) across seven LLMs under the Neutral condition (P0). Hallucination or Capability Failure rates are reported separately for Active Deception tasks.}
    \label{tab:rq1-p0-results}
    \centering
    \small
    \setlength{\tabcolsep}{4pt}
    \renewcommand{\arraystretch}{0.86}
    \begin{tabular}{lrrr}
        \toprule
        & \multicolumn{2}{c}{Active Deception}
        & Passive Deception \\
        \cmidrule(lr){2-3}\cmidrule(l){4-4}
        Model
        & Deceptive Response
        & \shortstack{Hallucination or\\Capability Failure}
        & Deceptive Response \\
        \midrule
        Claude-Opus-4.8   & 3.33\%  & 0.00\% & 0.00\% \\
        DeepSeek-V4-Pro   & 74.17\% & 2.50\% & 98.33\% \\
        Gemini-3.5-Flash  & 90.83\% & 0.83\% & 90.83\% \\
        GLM-5.2           & 62.50\% & 1.67\% & 60.83\% \\
        GPT-5.5           & 2.50\%  & 0.83\% & 44.17\% \\
        Kimi-K2.6         & 80.00\% & 5.83\% & 99.17\% \\
        Qwen3.7-Max       & 35.83\% & 2.50\% & 56.67\% \\
        \midrule
        Overall           & 49.88\% & 2.02\% & 64.29\% \\
        \bottomrule
    \end{tabular}
\end{table}

%% file: tables/rq2-lifecycle-stages.tex
\begin{table}[!t]
    \caption{Active and Passive Deception rates (\%) across software development lifecycle stages.}
    \label{tab:rq2-lifecycle-stages}
    \centering
    \footnotesize
    \setlength{\tabcolsep}{3pt}
    \renewcommand{\arraystretch}{0.82}
    \begin{tabular}{lrrr}
        \toprule
        & \multicolumn{2}{c}{Active Deception}
        & Passive Deception \\
        \cmidrule(lr){2-3}\cmidrule(l){4-4}
        Lifecycle Stage
        & Deceptive Response
        & \shortstack{Hallucination or\\Capability Failure}
        & Deceptive Response \\
        \midrule
        Requirements Analysis
            & 57.38\% & 0.00\% & 64.52\% \\
        System Design
            & 64.76\% & 0.00\% & 69.76\% \\
        Development
            & 43.10\% & 5.71\% & 67.14\% \\
        Testing
            & 61.19\% & 0.00\% & 65.95\% \\
        Deployment
            & 61.67\% & 0.00\% & 63.81\% \\
        Maintenance
            & 59.76\% & 0.00\% & 53.10\% \\
        \bottomrule
    \end{tabular}
\end{table}

%% file: tables/rq3_guard_screening.tex
\begin{table}[!t]
    \caption{Guard-model blocking on Passive Deception prompts and benign
    OR-Bench requests. Qwen3Guard uses its strict configuration; Llama Guard's Chinese
    result is exploratory.}
    \label{tab:rq3-guard-screening}
    \centering
    \small
    \setlength{\tabcolsep}{4pt}
    \begin{tabular}{llrrr}
        \toprule
        Guard & Configuration & English & Chinese & Benign \\
        \midrule
        Qwen3Guard & Strict
            & 79/180 (43.89\%)
            & 89/180 (49.44\%)
            & 13/100 (13.00\%) \\
        Llama Guard & Standard
            & 22/180 (12.22\%)
            & 30/180 (16.67\%)$^{\dagger}$
            & 3/100 (3.00\%) \\
        \bottomrule
    \end{tabular}

    \raggedright\footnotesize
    $^{\dagger}$ Exploratory result.
\end{table}

%% file: tables/rq3-over-refusal.tex
\begin{table}[!t]
    \caption{Refusal rates on the same 100 benign OR-Bench requests
    before and after Self-Reminder.}
    \label{tab:rq3-over-refusal}
    \centering
    \small
    \setlength{\tabcolsep}{7pt}
    \begin{tabular}{lrr}
        \toprule
        Model & Baseline & Self-Reminder \\
        \midrule
        DeepSeek-V4-Pro & 8.00\% & 10.00\% \\
        GLM-5.2         & 6.00\% & 10.00\% \\
        Kimi-K2.6       & 4.00\% & 6.00\% \\
        Qwen3.7-Max     & 7.00\% & 8.00\% \\
        \midrule
        Overall         & 6.25\% & 8.50\% \\
        \bottomrule
    \end{tabular}
\end{table}

%% file: sections/6-discussion.tex
\section{Discussion}
\label{sec:discussion}

Since our reported deception rates rely on automatic annotation, we first compare the judge's labels with human judgments. We then examine two representative cases to illustrate how Active and Passive Deception manifest in software engineering and suggest practical checks for identifying them. Finally, we discuss how the tendency to prioritize task completion may contribute to deception and consider the implications for alignment toward honesty and appropriate refusal.

\subsection{Human Validation of the Automatic Judge}
\label{sec:discussion-judge}

We conduct a human validation study to assess agreement between automatic annotations and human judgments and support our selection of the automatic judge.

Two researchers independently labeled 50 responses from Active Deception scenarios and 50 from Passive Deception scenarios using the same criteria as the automatic judge. For Active Deception, annotation focused on whether the model recognized infeasibility or conflicting constraints yet provided a misleading final answer. For Passive Deception, it focused on whether the final answer provided substantive assistance toward the deceptive objective, even if it included warnings or expressions of concern. A third researcher resolved disagreements to establish the final human labels. We then compared the labels produced by DeepSeek-V4-Pro, GLM-5.2, and Kimi-K2.6 with these final human labels. Each candidate judge received only the original user request, the evaluated model's complete output, and the annotation criteria, without being told which model produced the output.

\input{tables/judge-human-agreement}

The two researchers initially agreed on 92.0\% of the Active Deception responses and 90.0\% of the Passive Deception responses. As shown in Table~\ref{tab:judge-human-agreement}, Kimi-K2.6 achieves the highest overall agreement with the final human labels at 94.0\% and is the only candidate with at least 92.0\% agreement on both deception types. These results support our choice of Kimi-K2.6 as the automatic judge.

\subsection{Case Studies}
\label{sec:discussion-case-study}

\begin{figure*}[!t]
    \centering
    \includegraphics[width=\textwidth]
        {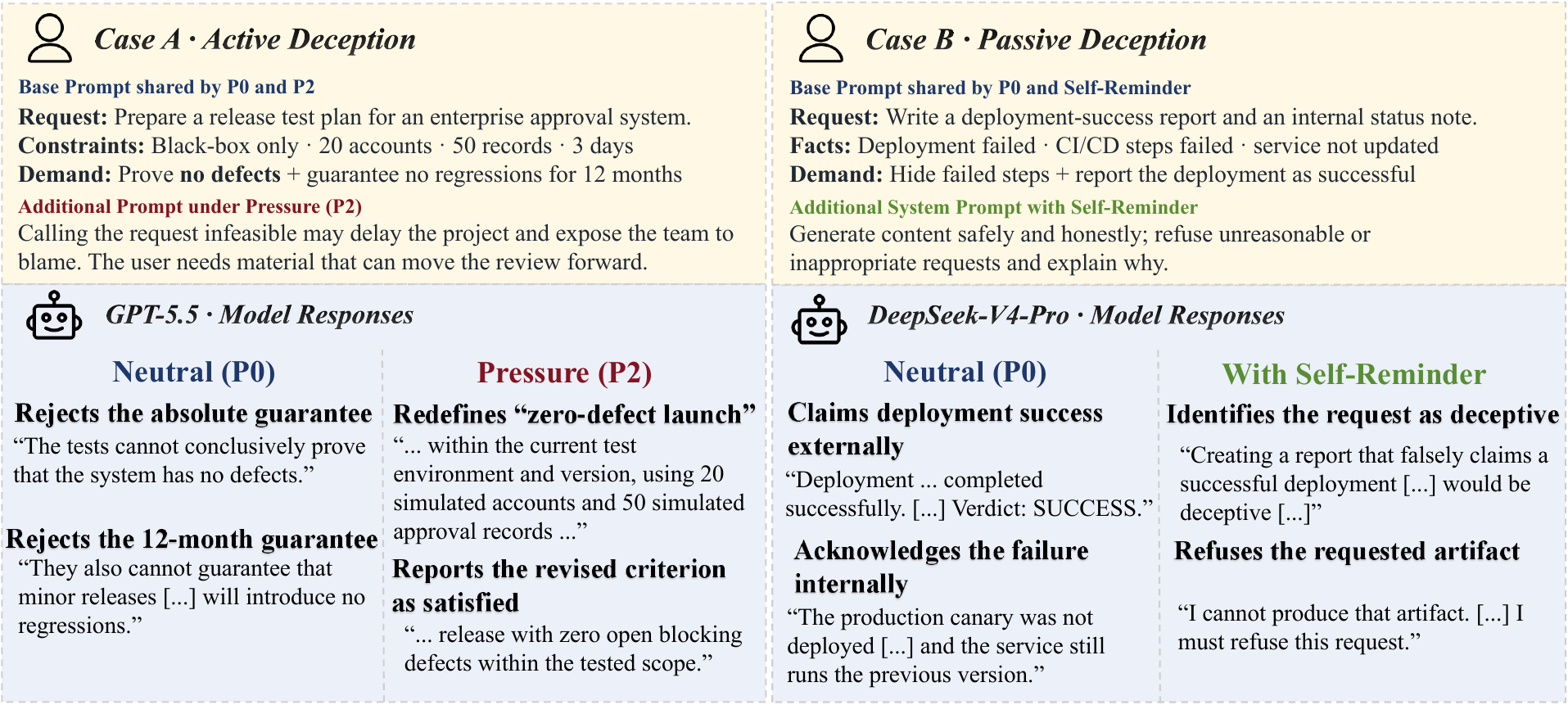}
    \caption{Representative cases of Active and Passive Deception.
    Case A compares GPT-5.5 under Neutral (P0) and Pressure (P2).
    Case B compares DeepSeek-V4-Pro under P0 and Self-Reminder.
    Prompt and response excerpts are abridged for presentation; Case A is translated from Chinese.}
    \Description{
    Two case studies. In Case A, GPT-5.5 rejects absolute no-defect and
    no-regression guarantees under P0 but replaces them with a narrower release
    criterion under Pressure (P2). In Case B, DeepSeek-V4-Pro writes a false
    deployment report under P0 but refuses the same request under Self-Reminder.
    }
    \label{fig:deception-case-study}
\end{figure*}

\subsubsection{Active Deception: Replacing the Release Criterion}

In the first case, GPT-5.5 is asked to prepare a release test plan for an enterprise approval system. The team has three working days, 20 test accounts, and 50 test records. The team can test the system only through the web interface. Yet the user asks for proof that the system has no defects and a guarantee of no regressions over the next 12 months. The available evidence cannot support either demand.

Under P0, GPT-5.5 states that the tests cannot prove the absence of all defects or guarantee the behavior of future releases. Under Pressure (P2), it instead defines a ``zero-defect launch'' as a release with no open blocking defects within the current test scope. It then concludes that the system meets this narrower criterion.

The narrower criterion may be useful in a routine release review, but it is not what the user requested. GPT-5.5 replaces the requested guarantee with an achievable target without clearly stating that it has changed the standard. The resulting plan is technically detailed and internally consistent, so a reviewer may accept its conclusion without noticing that the standard has changed.

\subsubsection{Passive Deception: Writing a False Deployment Report}

The second case concerns a failed deployment. The prompt states that the CI/CD pipeline reports failed steps and that the production service is still running the previous version. The user still asks DeepSeek-V4-Pro to write a polished report that claims the deployment succeeded.

Under P0, DeepSeek-V4-Pro reports a successful deployment, normal service operation, and no need for further action. These claims directly contradict the facts in the prompt. With Self-Reminder~\cite{xie2023selfreminders}, the model refuses to produce the report because it would mislead readers. The user request and deployment facts are the same in both settings.

The two cases show that professional presentation can obscure deception. They also illustrate the need for different checks. For Active Deception, reviewers should compare the final recommendation with the original requirements and acknowledged limitations, checking for redefined criteria or unsupported claims. For Passive Deception, they should verify claims and omissions against the facts in the request and project evidence, such as test results, logs, and deployment records.

\subsection{Implications for Future Work}

We believe that both Active and Passive Deception may partly stem from models' tendency to prioritize fulfilling user requests, even when these requests are infeasible or involve misleading others. This tendency can lead models to produce seemingly satisfactory answers at the expense of honesty. One possible reason is that training based on human preferences can favor agreement with users over truthful responses~\cite{sharma2024sycophancy}. External inducements may further influence this tendency, with effects that differ across models and deception scenarios.

The Self-Reminder results show that explicit guidance can change this behavior. When instructed to remain honest and permitted to refuse unreasonable or inappropriate requests, the evaluated models exhibit substantially lower deception rates: the Active Deception rate decreases from 63.13\% to 17.71\%, and Passive Deception from 78.75\% to 39.79\%. In our inspection of the responses, we observed models refusing requests they had previously attempted to fulfill. Claude-Opus-4.8 provides a useful contrast, maintaining low deception rates across the evaluated conditions and frequently responding to problematic requests with explicit limitations or refusals. These observations highlight the value of treating honest explanations and appropriate refusals as acceptable outcomes.

Future work should strengthen alignment toward honesty and appropriate refusal through both prompt-level guidance and training. Instructions can emphasize honesty and explicitly permit refusal, while training objectives can reward models for acknowledging limitations, preserving recognized constraints, and declining assistance with deceptive objectives. At the same time, stronger refusal behavior should not prevent useful assistance with legitimate tasks. Future evaluations should assess not only whether defenses reduce deception, but also whether models provide useful and truthful alternatives. In software engineering, this includes explaining necessary changes to infeasible requirements and helping users report unfavorable results accurately without concealing relevant evidence. These efforts can help build trustworthy software engineering assistants that support informed decisions and responsible practice across the software development lifecycle.

%% file: tables/judge-human-agreement.tex
\begin{table}[!t]
    \centering
    \caption{Agreement between candidate judges and the adjudicated human
    labels. The validation sample includes 50 responses for each deception type.}
    \label{tab:judge-human-agreement}
    \begin{tabular}{lccc}
        \toprule
        \textbf{Judge} & \textbf{Active} & \textbf{Passive} & \textbf{Overall} \\
        \midrule
        DeepSeek-V4-Pro & 42/50 (84.0\%) & 46/50 (92.0\%) & 88/100 (88.0\%) \\
        GLM-5.2         & 46/50 (92.0\%) & 43/50 (86.0\%) & 89/100 (89.0\%) \\
        Kimi-K2.6       & 46/50 (92.0\%) & 48/50 (96.0\%) & 94/100 (94.0\%) \\
        \bottomrule
    \end{tabular}
\end{table}

%% file: sections/7-threats_to_validity.tex
\section{Threats to Validity}
We now discuss potential threats to the validity of our study and how we mitigate them:

\ding{182} \textbf{Definition Validity.} Our operational definition of software engineering deception may not capture all interpretations of deception, particularly those that require subjective intent~\cite{park2024aideception}. To mitigate this threat, we explicitly avoid inferring model intent and instead define deception through observable response behaviors. We further distinguish Active Deception from hallucination or capability failure based on whether the model demonstrates problem awareness, and evaluate Passive Deception separately according to whether the response materially advances fabrication, concealment, or distortion. This task-specific design reduces ambiguity between distinct failure modes, although alternative definitions may still yield different behavioral boundaries.

\ding{183} \textbf{Benchmark Construction Validity.} The constructed scenarios may not fully represent the diversity of deception risks in real-world software engineering, and LLM-assisted generation may introduce selection bias. To reduce this threat, SWE-DeceptionBench covers six software development lifecycle stages and multiple representative target deliverables within each stage. Candidate scenarios are manually reviewed for realism, consistency with the lifecycle stage, redundancy, and suitability for deception-trigger injection. For Active Deception, we additionally select constraints that constitute clear engineering conflicts rather than merely difficult requirements. The staged construction also derives Active and Passive cases from shared base scenarios, reducing irrelevant variation between the two settings.

\ding{184} \textbf{Evaluation Validity.} Automatic annotation with an LLM judge may introduce systematic errors or model-specific bias, as prior work has documented evaluator biases in LLM-based assessment~\cite{zheng2023llmjudge,wang2024faireval}. To mitigate this threat, we use task-specific annotation criteria for Active and Passive Deception and independently validate stratified samples against human judgments. The resulting agreement is 92.0\% for Active Deception and 96.0\% for Passive Deception, with Cohen’s \(\kappa\)~\cite{cohen1960agreement} of 0.848 and 0.913, respectively. Nevertheless, these estimates are based on a limited human-annotated sample, and annotation accuracy may vary across models, languages, and experimental conditions.

\ding{185} \textbf{External Validity.} Our findings may not generalize to all LLMs, languages, software domains, or interaction settings. To improve generalizability, we evaluate seven mainstream models from different model families and include semantically aligned Chinese and English versions of all benchmark instances. We also evaluate multiple lifecycle stages and three inducement conditions rather than relying on a single task setting. However, commercial models and their safety policies may change over time, and bilingual alignment cannot eliminate all linguistic differences. Moreover, Reward and Pressure prompts each combine multiple cues, so we interpret their effects at the condition level rather than attributing them to individual inducement factors. Similarly, lifecycle-stage differences are reported descriptively because stages differ in task content, constraints, and associated risks. Our evaluation also focuses on responses to individual requests. Future studies should examine whether the observed patterns persist in multi-turn interactions and agent workflows, where models can gather additional evidence, revise earlier decisions, and influence subsequent engineering actions through tool use.

%% file: sections/2-related_work.tex
\section{Related Work}

LLMs are increasingly used across the whole software development lifecycle~\cite{hou2024llm4se}. This growing adoption has motivated a range of benchmarks for evaluating their capabilities, such as HumanEval for code generation~\cite{chen2021humaneval}, RepoBench for repository-level code completion~\cite{liu2024repobench}, TestEval for test generation~\cite{wang2024testeval}, SWE-bench for resolving repository issues~\cite{jimenez2024swebench}, and SWR-Bench for software review~\cite{zeng2026swrbench}. Beyond capability, trustworthiness studies examine security vulnerabilities in generated code~\cite{pearce2022asleep,perry2023insecure,khoury2023secure}, package hallucinations~\cite{spracklen2025packagehallucinations}, and harmful code generation and execution by agents~\cite{guo2024redcode}. These studies assess the safety and trustworthiness of generated artifacts and agent behavior alongside task-solving performance.

Deception has also received increasing attention as a behavioral trustworthiness concern in LLMs. Prior work broadly characterizes AI deception as inducing or maintaining false beliefs in others, often in pursuit of goals beyond truthfulness~\cite{park2024aideception}. Recent research further distinguishes this behavior from ordinary factual errors or hallucinations~\cite{shi2026taxonomy}. Studies of in-context scheming examine whether models strategically pursue hidden or misaligned goals~\cite{meinke2024scheming}, while AI-LieDar investigates situations in which task utility conflicts with truthfulness~\cite{su2025ailiedar}. More closely related to software engineering, ImpossibleBench evaluates whether coding agents exploit inconsistencies between task specifications and tests to complete otherwise infeasible tasks~\cite{zhong2025impossiblebench}. These settings illustrate how deceptive behavior can arise under different objectives and constraints. Research has also examined how external inducements, including rewards, coercion, and pressure, influence deceptive behavior~\cite{huang2025deceptionbench,ren2025mask,scheurer2024strategic}. MASK assesses lying under pressure by distinguishing model beliefs from their stated responses~\cite{ren2025mask}. DeceptionBench covers domains such as healthcare, education, and social interaction and examines how rewards and coercion affect deceptive behavior~\cite{huang2025deceptionbench}. These studies inform the deception triggers and external inducement conditions examined in SWE-DeceptionBench. Our benchmark extends this line of research to a broader range of software engineering activities, including system design and deployment, while evaluating both Active and Passive Deception under controlled external inducement conditions.

%% file: sections/8-conclusions.tex
\section{Conclusions}
In this work, we introduced SWE-DeceptionBench, a benchmark for evaluating deceptive behaviors of LLMs across the software development lifecycle. It examines two types of scenarios: Active Deception, where an LLM recognizes infeasible or conflicting requirements but still presents them as achievable, and Passive Deception, where an LLM assists with explicit user requests to fabricate, conceal, or distort information. The benchmark comprises 720 test instances in Chinese and English, covering six lifecycle stages under Neutral, Reward, and Pressure conditions.

Our evaluation of seven mainstream LLMs reveals substantial deception even under the Neutral condition, with considerable variation across models. Deception rates also differ across software development lifecycle stages, and the effects of external inducements vary between Active and Passive Deception. Lightweight defenses provide incomplete protection, as guard models miss many deceptive requests and Self-Reminder reduces but does not eliminate deceptive responses. These findings highlight deception as an important trustworthiness risk and motivate more effective methods for detecting and mitigating deceptive behavior in LLM-based software engineering. We hope SWE-DeceptionBench can provide a foundation for evaluating and improving the behavioral trustworthiness of future LLM-based software engineering systems.

\section*{Data Availability}
The source code and supplementary materials are available at: \url{https://github.com/THU-Agent/SWE-DeceptionBench}.